# QB-II for Evaluating the Reliability of Binary-State Networks

Wei-Chang Yeh
Integration and Collaboration Laboratory
Department of Industrial Engineering and Engineering Management
National Tsing Hua University
yeh@ieee.org
+886-986555381

*Abstract* — Current real-life applications of various networks such as utility (gas, water, electric, 4G/5G) networks, the Internet of Things, social networks, and supply chains. Reliability is one of the most popular tools for evaluating network performance. The fundamental structure of these networks is a binary state network. Distinctive methods have been proposed to efficiently assess binary-state network reliability. A new algorithm called QB-II (quick binary-addition tree algorithm II) is proposed to improve the efficiency of quick BAT, which is based on BAT and outperforms many algorithms. The proposed QB-II implements the shortest minimum cuts (MCs) to separate the entire BAT into main-BAT and sub-BATs, and the source-target matrix convolution products to connect these subgraphs intelligently to improve the efficiency. Twenty benchmark problems were used to validate the performance of the QB-II.

## 1. INTRODUCTION

Given the rapid development of advanced technologies, various networks have sprung up and become more diverse and practical, for example, some well-known networks include and traditional utility networks (e.g., water, gas, electricity, and telephony), Internet of Things [1], 4G/5G telecommunications [2], social networks [3], deep learning [4, 5], cloud/fog/edge computing [6] and smart wireless sensor networks [7, 8]. Numerical networks have become a part of everyday life for nearly every human and industry (e.g., manufacturing, commerce, and supply chains) worldwide [9, 10, 11].

All types of networks [12, 13, 14, 15, 16, 17] are derived from a binary-state network in which the state of each component is binary: success or failure. For example, the multi-state flow network is a special binary-state network, but each component has multiple states [18, 19, 20]. The information network is a special multi-state network without satisfying the conservation law [11, 21]. The multi-commodity multi-state flow/information network is a special multi-state flow/information network that allows different types of flows [22, 23]. The multi-distribution multi-commodity multi-state flow/information network is a special multi-commodity multi-state flow/information network, such that each component has more than one probability distribution [24]. Hence, the improvement of the algorithm in calculating the binary-state network reliability can help advance reliability assessment in other types of networks [20, 21, 22, 23, 24].

Owing to the importance and practice of all types of networks, we need to have an index to evaluate the performance of networks for a better understanding and management of networks. Reliability is a popular index for evaluating network performance. Reliability is the success probability that the network is still functioning under pregiven conditions, such as flow amount [20, 21, 22, 23, 24], signal quality [25], budget [13, 18, 26], transition time [15, 27, 28], and production numbers [29].

It is both NP-hard and #P-hard to calculate the reliability of binary-state networks [9, 10, 11]. There are three categories in traditional network reliability calculation (e.g., transportation and communication): exact-reliability algorithms [30, 31], approximated-reliability algorithms [32, 33, 34], and the combination of both [35, 36]. The major factor among these categories is whether all connected and disconnected vectors are needed to calculate the network reliability and network unreliability, respectively.

The exact-reliability algorithms need to find all feasible related special vectors, such as the minimal cuts (MCs) in MC-algorithms [20, 28, 37, 38], the minimal paths (MPs) in MP-algorithms [13, 18, 24, 26, 27, 29, 30, 31], the connected vectors binary-addition-tree algorithms (BATs) [39, 40, 41] and the binary decision diagram (BDD) [40, 42], to have the exact reliability. The sizes of the special MCs and MPs are smaller than those of the connected vectors. However, after having all MCs or all MPs, MC-algorithms and MP-algorithms must solve another NP-hard problem using inclusion-exclusion techniques (IETs) [43, 44] or sum-of-disjoint product methods (SDPs) [30, 45, 46] in terms of the identified MPs or MCs to have the final exact reliability. Hence, these direct and straightforward methods, for example, BATs [39, 40, 41] and BDD [40, 42], without the need to have MPs and MCs, are becoming more popular than MC-algorithms and MP-algorithms.

Approximated-reliability algorithms rely only on parts of the connected vectors, for example, Monte Carlo simulations (MCSs) [32, 36], bounds [34, 35], and AI [4, 5, 33, 47, 48], to obtain approximate reliability quickly. However, we do not know how well the approximated reliabilities are. For MCS and AI, there is no confirmation that the obtained approximated reliability is larger or less than the exact reliability [32, 33, 36, 48]. The bound proposed in the bound algorithms is either too far away from the exact reliability [34, 35] if only a few connected vectors are used or still requires too much time to obtain if most of the connected vectors must be found in advance.

The methods combining both exact- and approximated-reliability have the advantages of both exact and approximated reliability, and may be the future of the research area of network reliabilities to have a better approximated reliability than the approximated-reliability algorithm and more efficiency than the exact-reliability algorithms [36].

From the above discussion, the exact-reliability algorithm takes more time than the other two categories in calculating reliability. The exact-reliability algorithm is not suitable for these complex networks if the time is more important than the accuracy to have the reliability estimate. However, the advances in new algorithms and computer abilities in dealing with complicated calculations have made the exact-reliability algorithms more practical and efficient than before.

For example, in [41], the BAT (the exact-reliability algorithm) can solve the benchmarks for the redundancy allocation problem within one minute, which has been impossible for more than 54 years [49]. Hence, there is always a need to have a better exact-reliability algorithm.

Yeh first proposed BAT to generate all possible binary vectors [39]. The quick BAT, also proposed by Yeh, is an improved BAT [39, 40] that can filter out all connected vectors without completely removing each infeasible vector. The quick BAT is also a direct algorithm for calculating network reliability without using MPs and MCs. It integrates the first connected vector, the last disconnected vector, and super vectors, and the corresponding probability calculations that make it faster than traditional methods such as the BDD [40, 42] and the BAT, which outperforms DFS, BFS, and UGFM [39, 40].

The main purpose of this research was to propose a new quick BAT called QB-II to improve the efficiency of the quick BAT in calculating the binary-state reliability. The proposed QB-II implements the shortest minimum cuts (MCs) to separate the entire BAT into the main BAT and sub-BATs. QB-II adapted a new convolution product called the source-target matrix convolution product to connect these BATs intelligently to reduce both the number of calculations of probabilities and the verifications of connections.

The remainder of this paper is organized as follows. All the required acronyms, notations, nomenclature, and assumptions needed in the proposed QB-II are presented in Section 2. Section 3 reviews the fundamentals of fundamental graph theory, quick BAT, and traditional convolution products. Section 4 introduces the novel concepts in separating the graph into subgraphs together with several novel concepts: the shortest MCs, sub-BATs, aggregated source-target matrices, and their probabilities. Section 5 provides a novel convolution product based on source-target matrices to improve the efficiency of QB-II. Section 6 presents the QB-II pseudo-code, a demonstration, and experimental results. Finally, Section 7 concludes the paper.

## 2. ACRONYMS, NOTATIONS, NOMENCLATURE, AND ASSUMPTIONS

All required acronyms, notations, nomenclature, and assumptions are provided in this section.

## 2.1 Acronyms

   MC/MP: minimal cut/path

  min-cut: minimum cut

    BAT: binary-addition tree algorithm [39]

  QB-II: proposed new quick BAT [40]

   BDD: binary decision diagram

   BFS: breadth-first-search method

   DFS: depth-first-search method

UGFM: universal generating function methodology

   IET: inclusion–exclusion technology

   QIE: quick IET [44]

   SDP: sum-of-disjoint products method

## 2.2 Notations

  $|\bullet|$: number of elements in $\bullet$

 Pr($\bullet$): success probability of event $\bullet$

    $n$: number of nodes

    $m$: number of arcs

    $V$: set of nodes $V = \{1, 2, \ldots, n\}$

    $E$: set of arcs $E = \{a_1, a_2, \ldots, a_m\}$

   $a_k$: $k^{\text{th}}$ arc in $E$

   $\mathbf{D_b}$: The state distribution lists the functioning probability for each arc, e.g., Table 1 is an state distribution of all arcs in Figure 1.

**Table 1.** Example of arc state distribution.

| $i$ | Pr($a_i$) | $i$ | Pr($a_i$) |
|---|---|---|---|
| 1 | 0.98 | 5 | 0.75 |
| 2 | 0.80 | 6 | 0.90 |
| 3 | 0.85 | 7 | 0.88 |
| 4 | 0.95 | | |

$G(V, E)$: A graph with $V$, $E$, source node 1, and a sink node $n$, for example, Figure 1 is a graph with $V = \{1, 2, 3, 4, 5\}$, $E = \{a_i \mid i = 1, 2, \ldots, 7\}$, source node 1, and sink node 4.

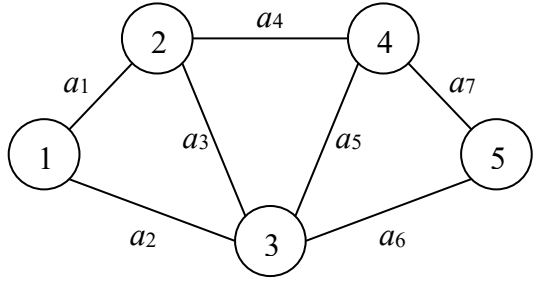

**Figure 1.** Example network.

$G(V, E, \mathbf{D_b})$: A binary-state network with $G(V, E)$ and $\mathbf{D_b}$, e.g., Figure 1 is a binary-state network $G(V, E, \mathbf{D_b})$ if $\mathbf{D_b}$ is presented in Table 1.

$\otimes$: Convolution product

$X$: (state) vector

$X_{FC}$: first connected (state) vector in the proposed bound BAT

$X_{LD}$: last disconnected (state) vector in the proposed bound BAT

$X(i)$: state of the $i^{th}$ coordinate in $X$ for $i = 1, 2, \ldots, |X|$

$X(a_i)$: state of $a_i$ in $X$ for $i = 1, 2, \ldots, |X|$

$\Pr(X(a_i))$: $\Pr(X(a_i)) = \Pr(a_i)$ and 0 if $\Pr(X(a_i)) = 1$ and 0 for $i = 1, 2, \ldots, |X|$, respectively

$\Pr(X)$: $\Pr(X) = \sum_{i=1}^{|X|} \Pr(X(a_i))$

$S_{SV}(X_S)$: the vector subset related to the super vector $X_S$.

$G(X)$: $G(X) = G(V, E(X))$ and $E(X) = \{ a_i \mid X(a_i)=1 \text{ for } i = 1, 2, \ldots, |X|\}$, e.g., $X = (0, 1, 1, 1, 0, 1, 1)$ as shown in Figure 2 and $G(X)$ is a subgraph of that in Figure 1.

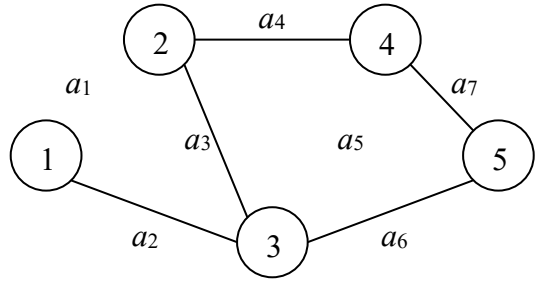

**Figure 2.** $G(X)$ and $X = (0, 1, 1, 1, 0, 1, 1)$ in Figure 1.

$R(G)$: the network reliability.

$R(G(X))$: reliability of $G(V, E(X), \mathbf{D_b})$

$n_p$: The number of arcs in the shortest paths from nodes 1 to $n$, e.g., $\{a_2, a_6\}$ is the shortest path in Figure 1 and $n_p = 2$.

$n_c$: The number of arcs in any minimum cut between node 1 and node $n$, e.g., $\{a_1, a_2\}$ and $\{a_6, a_7\}$ are the minimum cuts in Figure 1 and $n_c = 2$.

$A \leq B$: $A(a_i) \leq B(a_i)$ for all $i = 1, 2, \ldots, m$, e.g., $(1, 1, 1, 0, 1) \leq (1, 1, 1, 1, 1)$

$A < B$: $A \leq B$ and $A(a_i) < B(a_i)$ for at least one $j = 1, 2, \ldots, m$. For example, $(1, 1, 1, 0, 1) < (1, 1, 1, 1, 1)$

$A \ll B$: vector $A$ is obtained after $B$ in the BAT, e.g., $(0, 1, 0, 1, 1) \ll (1, 1, 0, 1, 1)$.

$A \underline{\ll} B$: $A \ll B$ or $A = B$.

## 2.3 Nomenclature

| | |
|---|---|
| Network Reliability: | The success probability that node 1 is connected to node $n$ in network. |
| Shortest Path: | A path from nodes 1 to $n$ such that the number of arcs is minimized in such path. |
| Minimum Cut: | A cut between nodes 1 to $n$ such that the number of arcs is minimized in such cut. |
| Connected vector: | Vector $X$ is connected if nodes 1 and $n$ are connected in $G(X)$ [39]. |
| Disconnected vector: | Vector $X$ is disconnected if nodes 1 and $n$ are disconnected in $G(X)$ [39]. |

First connected vector: The connected vector such that any vector obtained before it is always disconnected in the BAT [40].

Last disconnected vector: The disconnected vector such that any vector obtained after it is always connected in the BAT [40].

### 2.4 Assumptions

1. All nodes are perfectly reliable and connected.

2. Each arc state is either functioning or has failed.

3. There is no parallel arcs or loops.

4. The probability of each arc is statistically independent with a pregiven distribution.

### 3. CUTS, PATHS, QUICK BAT AND CONVOLUTION PRODUCT

The proposed QB-II is an improved quick BAT [40] with a new concept that separates the white networks into smaller networks using the minimum cuts (min-cuts) and one shortest path and integrates these vectors into subgraphs using the convolution product. Hence, the shortest paths, min-cuts, quick BAT, and convolution product are discussed in this section.

### 3.1 Shortest Paths and Minimum Cuts

The shortest path $P^*$ is the path from nodes 1 to $n$ and $W(P^*) \leq W(P)$ for all paths $P$. For example, $\{a_2, a_6\}$ is the only shortest path in Figure 1. Determining the shortest paths in graph theory is a classical problem. The Dijkstra algorithm was adapted in the proposed algorithm because it requires only $O(|E|)$ to find the shortest path [51].

All cuts are special arc subsets, and their removal disconnects nodes 1 and $n$. A minimum-cut (min-cut) $C^*$ is a special cut such that $W(C^*) \leq W(C)$ for all cuts $C$, that is, the total weight of arcs in any min-cut is a minimum. For instance, in Figure 1, there are two min-cuts: $\{a_1, a_2\}$ and $\{a_6, a_7\}$.

The Stoer-Wagner algorithm is one of the most efficient in finding all min-cuts in $O(mn + n^2 \log$

*n*) [52], and it was adapted in the proposed QB-II to separate the binary-state network.

## 3.2 Quick BAT

The quick BAT proposed in [40] is an improved BAT. It outperforms traditional binary-state reliability algorithms such as BDD, DFS, BFS, UGF, and BAT [39, 40]. The major innovations include three special vectors: the first connected vector $X_{FC}$, the last disconnected vector $X_{LD}$, and the connected super vector. Based on these three special vectors, the quick BAT only finds connected super vectors after $X_{FC}$ and before $X_{LD}$ rather than all connected vectors in the BAT, to improve the efficiency of the BAT. The details of these three vectors are briefly discussed here.

### 3.2.1 BAT

BAT, which was originally developed by Yeh [39], belongs to the family of heuristic algorithms and can generate all required binary vectors.

In BAT, the initial binary vector $X = 0$ is updated repeatedly based on the following two rules:

**Rule 1.** The first zero-value coordinate $a_i$ is changed to one, and $a_j = 0$ for $j < i$.

**Rule 2.** If no such coordinate exists in Rule 1, halt and all vectors are found.

For example, (0, 0, 1) and (1, 0, 1) are updated to (1, 0, 1) and (0, 1, 1), respectively, and the entire process halts if the vector is (1, 1, 1), which has no zero-value coordinates.

BAT presents some difficulties in comprehension, coding, and customization [39, 40]. Hence, BAT has been applied to many research areas such as the resilience assess [53], the spread of wildfires [54], propagation of computer viruses [55], and the reliability of various networks [36, 39, 40, 41, 48].

The pseudo-code of the BAT is carried as follows [56]:

**Algorithm for finding All *m*-tuple Binary-State Vectors**

**Input:** *m*.

**Output:** All *m*-tuple binary-state vectors.

**STEP B0.** Let *X* be a vector zero and let $i = 1$.

**STEP B1.** If $X(a_i) = 0$, let $X(a_i) = 1$, $i = 1$, a new *X* is found, and go to STEP B1.

**STEP B2.** If $i = m$, halt.

**STEP B3.** Let $X(a_i) = 0$, $i = i + 1$, and go to STEP B1.

After obtaining all the connected vectors using the above BAT pseudocode, we can calculate the exact reliability using the following equation:

$$R(G) = \sum \Pr(X) = \sum \left[ \prod_{i=1}^{m} \Pr(X(a_i)) \right] \quad \text{for all connected vector } X. \tag{1}$$

For example, seven arcs are shown in Figure 1 and $2^7 = 128$. Hence, we have 128 different binary vectors after generating the BAT pseudo-code.

### 3.2.2 First Connected Vector

From the definition of the first connected vector, $X_{FC}$, *X* is disconnected if $X \ll X_{FC}$. Hence, there is no connected vector before $X_{FC}$ and we can find $X_{FC}$ at the beginning of BAT without needing to find any vector that is less than $X_{FC}$.

The basic idea in finding $X_{FC}$ was first proposed in quick BAT [40]. Hence, the algorithm used in the proposed algorithm was based on [40]. In [40], the $X_{FC}$ is obtained by letting the weight, called the FC weight here, of $a_i$ be $2^{m-i}$ for all $i = 1, 2, \ldots, m$ and finding the shortest path from nodes 1 to *n*. Then, $X_{FC}(a_i) = 1$ if $a_i$ is in the shortest path; otherwise, $X_{FC}(a_i) = 0$ for all $i = 1, 2, \ldots, m$.

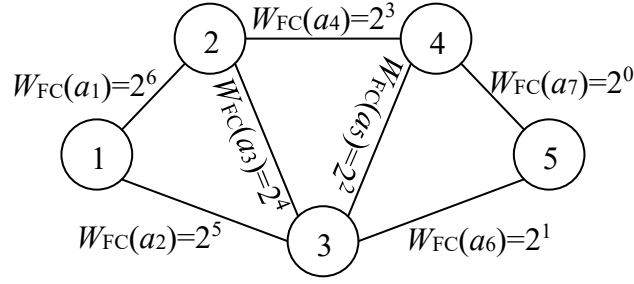

**Figure 3.** Example graph after having FC weight of each arc in Figure 1.

For example, the FC-weight of all arcs in Figure 1 are set as in Figure 2 by letting $W(a_i) = 2^{7-i}$ for $i = 1, 2, …, 7$. Because $P^* = \{a_2, a_6\}$ is the shortest path, we have $X_{FC} = (0, 1, 0, 0, 0, 1, 0)$. The BAT can search for new vectors from $X_{FC}$ not from vector zero, and it helps us to skip $1 \times 2^5 + 1 \times 2^1 = 34$ vectors from the traditional BAT in Figure 1.

### 3.2.3 Last Disconnected Vector and Its aggregated Probability

In contrast to the concept of the first connected vector $X_{FC}$, the last disconnected vector is the disconnected vector $X_{LD}$ such that any vector found after it is always connected in the BAT process. However, similar to $X_{FC}$, we can skip all vectors $X$ with $X_{LD} \ll X$ after finding $X_{LD}$, that is, halt the BAT immediately to improve its efficiency.

The original method for finding $X_{LD}$ is also proposed in quick BAT, and its main concept is to find the minimum cut (min-cut) between nodes 1 and $n$ after setting the weight called the LD weight $W(a_i) = 2^i$ for all $i = 1, 2, …, m$. After the min-cut, $X_{LD}(a_i) = 0$ and 1 if $a_i$ is and is not in the min-cut, respectively, for all $i = 1, 2, …, m$.

Both methods of setting the weight in finding $X_{FC}$ and $X_{LD}$ were the same. However, the shortest path is required for $X_{FC}$ and the min-cut is required for $X_{LD}$.

For example, each arc weight is shown in Figure 3 by setting $W(a_i) = 2^i$ for $i = 1, 2, …, 7$. There is only one min-cut, $C = \{a_1, a_2\}$, and $X_{LD} = (0, 0, 1, 1, 1, 1, 1)$. Hence, BAT can be halted after having $X_{LD}$ and the last three vectors, $(1, 0, 1, 1, 1, 1, 1)$, $(0, 1, 1, 1, 1, 1, 1)$, and $(1, 1, 1, 1, 1, 1, 1)$ are connected and skipped.

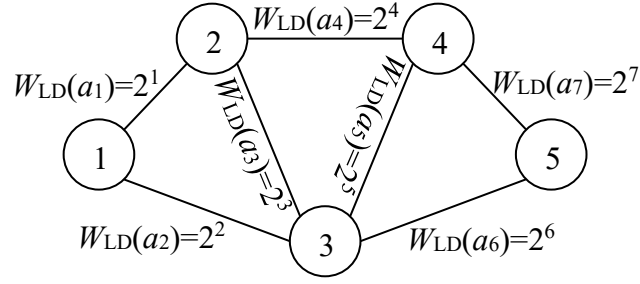

**Figure 4.** Example graph after having LD weight of each arc in Figure 1.

If $X_{LD} \ll X$, we have

$$\sum \Pr(X) = \sum \left[ \prod_{i=1}^{m} \Pr(X(a_i)) \right]$$

After all the connected vectors are obtained using the above BAT pseudocode, we can calculate the exact reliability using the following equation:

$$R(G) = \sum \Pr(X) = \sum \left[ \prod_{i=1}^{m} \Pr(X(a_i)) \right] \quad \text{for all connected vector } X. \qquad (2)$$

### 3.2.4 Super Vector and Its aggregated Probability

Based on [40], a super vector, for example, $X_s$ with $|X_s|$ tuples, is a special subvector such that the value of its $i$th coordinate is equal to that of the $i$th coordinate of the original vector $X$ for $i = 1, 2, \ldots, |X_s|$. Hence, $X_s$ can be a subvector of any $m$-tuple vector in $S(X_s) = \{X \mid X_s(a_i) = X(a_i) \text{ for } i = 1, 2, \ldots, |X_s|\}$ and $|S(X_s)| = 2^{m^*}$, where $m^* = m - |X_s|$. For example, in Figure 1, let $X_s = (1, 0, 1, 0, 0, 1)$, we have $S(X_s) = \{(1, 0, 1, 0, 0, 1, 0) \text{ and } (1, 0, 1, 0, 1, 1)\}$.

In addition, any $m$-tuple vector has $m$ sub-vectors. For example, (1, 0, 1, 0, 0, 1, 0) has eight super vectors: (1), (1, 0), (1, 0, 1), (1, 0, 1, 0), (1, 0, 1, 0, 0), (1, 0, 1, 0, 0), (1, 0, 1, 0, 0, 1), and (1, 0, 1, 0, 0, 1, 0).

The most important characteristic of the super vector is that any vector $X$ in $S(X_s) = X(a_i)$ is connected if $X_s$ is connected. Hence, there is no need to find any vector in $S(X_s)$ after knowing that $X_s$ is connected, and this can save time to find all vectors in $S(X_s)$.

For example, in Figure 1, $X_s = (1, 0, 1, 0, 0, 1)$. Because $X_s$ is connected, any vector in $S(X_s) = \{(1, 0, 1, 0, 0, 1, 0)$ and $(1, 0, 1, 0, 1, 1)\}$ is also connected, and there is no need to determine and verify the connectivity of any vector in $S(X_s)$.

The probability $\Pr(S(X_s))$ of the connected super vector $X_s$ for all vectors in $S(X_s)$ can be calculated in the way similar to that for the connected vector but ignores these $a_i$ for $i = |X_s|+1$, $|X_s|+2$, …, $m$. For example, let $X_s = (0, 1, 0, 1, 1)$ be a super vector and $\Pr(X(a_i)) = 0.8$ and $0.2$ if $X(a_i) = 1$ and $0$, respectively. We have $\Pr(S(X_s)) = 0.8^3 \times 0.2^2 = 0.02048$ [40].

Form the above, Eq. (2) can be simplified further after finding all the connected vectors:

$$R(G) = \sum \Pr(X) = \sum \left[ \prod_{i=1}^{|X_s|} \Pr(X_s(a_i)) \right] \tag{3}$$

where for all connected super vector $X_s$ with $X_{FC} \ll X_s \ll X_{LD}$.

## 3.3 Traditional Convolution Product

In the BAT proposed in [], after separating the problems into subproblems and finding all vectors for each subproblem, these vectors are integrated recursively using the convolution product $\otimes$ from one subproblem to the next subproblem.

Let $X_{i,j} = (x_{i,j,1}, x_{i,j,2}, …, x_{i,j,|Gi|})$ be the $j$th vector in the sub-BAT related to the $i$th subproblem $G_i$. The convolution product $\otimes$ is defined as follows for two vectors and two BATs, respectively:

$$X_{i,j} \otimes X_{(i+1),k} = (x_{i,j,1}, x_{i,j,2}, …, x_{i,j,|Gi|}, x_{(i+1),k,1}, x_{(i+1),k,2}, …, x_{(i+1),k,|G(i+1)|}), \tag{4}$$

$$\Xi_i \otimes \Xi_{(i+1)} = \{ X \mid X = X_{i,j} \otimes X_{(i+1),k} \text{ for } j = 1, 2, …, |\Xi_i| \text{ and } k = 1, 2, …, |\Xi_k|\}. \tag{5}$$

For example, the results of $X_1 \otimes X_2$ are listed in Table 2 for $X_1 \in \{(0, 0), (1, 0), (0, 1), (1, 1)\}$ and $X_2 \in \{(0, 0, 0), (1, 0, 0), (0, 1, 0), (1, 1, 0), (0, 0, 1), (1, 0, 1), (0, 1, 1), (1, 1, 1)\}$.

**Table 2.** Results for $X_1 \otimes X_2$.

| $i$ | $X_1$ | $X_2$ | $X_1 \otimes X_2$ | $i$ | $X_1$ | $X_2$ | $X_1 \otimes X_2$ |
|---|---|---|---|---|---|---|---|
| 1 | (0, 0) | (0, 0, 0) | (0, 0, 0, 0, 0) | 17 | (0, 1) | (0, 0, 0) | (0, 1, 0, 0, 0) |
| 2 | (0, 0) | (1, 0, 0) | (0, 0, 1, 0, 0) | 18 | (0, 1) | (1, 0, 0) | (0, 1, 1, 0, 0) |
| 3 | (0, 0) | (0, 1, 0) | (0, 0, 0, 1, 0) | 19 | (0, 1) | (0, 1, 0) | (0, 1, 0, 1, 0) |
| 4  | (0, 0) | (1, 1, 0) | (0, 0, 1, 1, 0) | 20 | (0, 1) | (1, 1, 0) | (0, 1, 1, 1, 0) |
| 5  | (0, 0) | (0, 0, 1) | (0, 0, 0, 0, 1) | 21 | (0, 1) | (0, 0, 1) | (0, 1, 0, 0, 1) |
| 6  | (0, 0) | (1, 0, 1) | (0, 0, 1, 0, 1) | 22 | (0, 1) | (1, 0, 1) | (0, 1, 1, 0, 1) |
| 7  | (0, 0) | (0, 1, 1) | (0, 0, 0, 1, 1) | 23 | (0, 1) | (0, 1, 1) | (0, 1, 0, 1, 1) |
| 8  | (0, 0) | (1, 1, 1) | (0, 0, 1, 1, 1) | 24 | (0, 1) | (1, 1, 1) | (0, 1, 1, 1, 1) |
| 9  | (1, 0) | (0, 0, 0) | (1, 0, 0, 0, 0) | 25 | (1, 1) | (0, 0, 0) | (1, 1, 0, 0, 0) |
| 10 | (1, 0) | (1, 0, 0) | (1, 0, 1, 0, 0) | 26 | (1, 1) | (1, 0, 0) | (1, 1, 1, 0, 0) |
| 11 | (1, 0) | (0, 1, 0) | (1, 0, 0, 1, 0) | 27 | (1, 1) | (0, 1, 0) | (1, 1, 0, 1, 0) |
| 12 | (1, 0) | (1, 1, 0) | (1, 0, 1, 1, 0) | 28 | (1, 1) | (1, 1, 0) | (1, 1, 1, 1, 0) |
| 13 | (1, 0) | (0, 0, 1) | (1, 0, 0, 0, 1) | 29 | (1, 1) | (0, 0, 1) | (1, 1, 0, 0, 1) |
| 14 | (1, 0) | (1, 0, 1) | (1, 0, 1, 0, 1) | 30 | (1, 1) | (1, 0, 1) | (1, 1, 1, 0, 1) |
| 15 | (1, 0) | (0, 1, 1) | (1, 0, 0, 1, 1) | 31 | (1, 1) | (0, 1, 1) | (1, 1, 0, 1, 1) |
| 16 | (1, 0) | (1, 1, 1) | (1, 0, 1, 1, 1) | 32 | (1, 1) | (1, 1, 1) | (1, 1, 1, 1, 1) |

Hence, all vectors, including connected vectors and disconnected vectors in $G$ can be found in $\bigotimes_{i=1}^{\eta} \Xi_i$, where $\eta$ is the number of subgraphs after using the shortest MCs. The benefit of using the convolution product $\otimes$ is that it discards infeasible vectors as soon as they are found. For example, there is no need to include $X_{i,1}$ in any vector using the convolution product $\otimes$ if $X_{j,1}$ is infeasible; that is, $\bigotimes_{i=1}^{\eta} X_i$ can be reduced to

$$\bigotimes_{i=1}^{\eta} X_i = \{ X \mid X = X_{1,k_1} \otimes X_{2,k_2} \otimes \ldots \otimes X_{\eta,k_\eta} \text{ for } k_j = 1, 2, \ldots, (i-1), (i+1), \ldots, \eta, \text{ and } j = 1, 2, \ldots, \eta\}. \tag{6}$$

## 4. SUBPROBLEMS

There are two major parts in QB-II, and these two parts are also the differences between the proposed QB-II and quick BAT:

1. QB-II separates the entire problem into smaller subproblems and finds and calculates the probability of each connected vector in smaller subproblems.
2. The QB-II combines these results from subproblems.

The details of two parts are discussed in Sections 4 and 5, respectively.

## 4.1 Shortest MCs

The first step of the proposed algorithm is to separate the graph into subgraphs. The core of separating the graph is based on $|P^*|$ parallel min-cuts, where $|P^*|$ is the number of arcs in the shortest path $P^*$.

The shortest path, $P^*$ can be easily found using the fundamental concept in books related to graph theory. In $|P^*|$ parallel min-cuts, each min-cut is generated based on $P^*$ such that each arc in $P$ is included in one and only one min-cut, and such $|P^*|$ parallel min-cuts are called the shortest MCs. The pseudocode of the proposed shortest MCs is given as follows:

**Algorithm for finding all shortest MCs**

**Input:**   A graph $G(V, E)$ with source node 1, sink node $n$, and the shortest path $P^*$.

**Output:**   All shortest MCs.

**STEP M0.**   Let $G^\# = G(V, E)$, $S = \{1\}$, and $i = j = 1$.

**STEP M1.**   Find a min-cut $C_i$ that separates $S$ and $n$ by splitting $G^\#$ into $G_i(S)$ and $G_i(T)$ such that the $i$th arc in $P^*$ is also in $(P^* \cap C_i) \subseteq G_i(S)$, $S_i \subseteq G_i(S)$, and $n \in G_i(T)$. If no such cut exists, proceed to STEP M3.

**STEP M2.**   Let $i = i + 1$, $S_i = \{v \mid \text{for all nodes } v \text{ in } C_i \cap G_{(i-1)}(T)\}$, and go to STEP M1.

**STEP M3.**   If $i < |P^*|$, let $i = i + 1$ and go to STEP M1. Otherwise, halt.

It takes $O(|V|^2)$ to find one of the shortest paths, say $P^*$, using the Dijkstra algorithm [51], and the number of arcs in $P^*$ is at most $|V|$. The time complexity is $O(|V| \times |E|)$ to find a min-cut using any maximum-flow algorithm, for example, Dinic's algorithm [51] in STEP M1. Note that a min-cut is the shortest MC, except that the shortest MC must include only one arc in $P^*$. Hence, the time complexity for finding all the shortest MCs is $O(|V|^2 + |V| \times |V| \times |E|) = O(|V|^2|E|)$.

For example, in Figure 1, $P^* = \{a_2, a_6\}$ is the shortest path from nodes 1 to 5, $C_1 = \{a_1, a_2\}$ and $C_2 = \{a_6, a_7\}$ are two parallel MCs with $P^* \cap C_1 = \{a_1\}$ and $P^* \cap C_2 = \{a_6\}$. Hence, $C_1 = \{a_1, a_2\}$ and $C_2 = \{a_6, a_7\}$ are the two shortest MCs, respectively. The information of $S$, $V(G_i(S))$, and $V(G_i(T))$ for $i = 1, 2$, and 3 is listed in Table 3, and three subgraphs are shown in Figure 5.

Table 3. Shortest MCs in Figure 1.

| $i$ | the $i$th arc in $P^*$ | $C_i$ | $S$ | $V(G_i(S))$ | $V(G_i(T))$ |
|---|---|---|---|---|---|
| 1 | $a_2$ | $\{a_1, a_2\}$ | $\{1\}$ | $\{1\}$ | $\{2, 3, 4, 5\}$ |
| 2 | $a_6$ | $\{a_6, a_7\}$ | $\{2, 3\}$ | $\{2, 3, 4\}$ | $\{5\}$ |

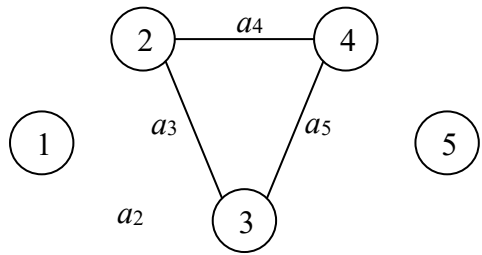

Figure 5. Subgraphs after separating by $C_1 = \{a_1, a_2\}$ and $C_2 = \{a_6, a_7\}$ in Figure 1.

### 4.2 Self-Adjustment Subgraphs

The arcs in each subgraph determine the size of the corresponding BAT. It is trivial that the arcs adjacent to two nodes in the same subgraph belong to the same BAT, for example, all arcs in $G_i(E)$. The problem is the arc between two consecutive subgraphs, for example, $C_i$. To prevent some subgraphs from having too many arcs and some having few arcs, let these arcs in $C_i$ be included in the BAT for $G_i$ if $G_i(E) + |C_i| < G_{(i-1)}(E)$; otherwise, $C_i$ is included in the BAT for $G_{(i+1)}$.

For example, in Figures 1 and 5, based on Table 3, we have a new arc subset for each subgraph listed in Table 4 and shown in Figure 6 (with different colored and dashed lines) after the self-adjustment.

Table 4. Adjusted subgraphs in Figure 1.

| $i$ | $C_i$ | Original $G(E_i)$ | Adjusted $G(E_i)$ |
|---|---|---|---|
| 1 | $\{a_1, a_2\}$ | ∅ | $\{a_1, a_2\}$ |
| 2 | $\{a_6, a_7\}$ | $\{a_3, a_4, a_5\}$ | $\{a_3, a_4, a_5\}$ |
| 3 | | ∅ | $\{a_6, a_7\}$ |

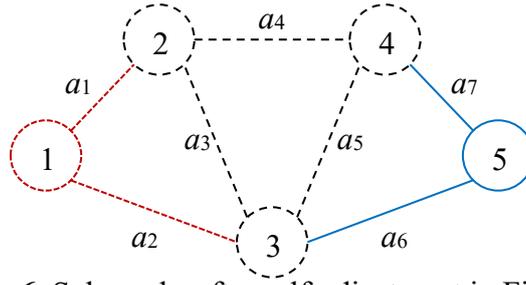
**Figure 6.** Subgraphs after self-adjustment in Figure 5.

### 4.3 Main-BAT and Sub-BATs

Let $B_{(i)}$ be the set of all $i$-tuple vectors, and $B_i$ be the set of all $|G_i(E)|$-tuple vectors related to the $i$th self-adjustment subgraph. Both $B_{(i)}$ and $B_i$ can be obtained from a traditional BAT in a straightforward manner. To increase the efficiency and save the computer memory, only $B_i$ needs to be found by implementing the BAT to the largest subproblems, that is, the subgraph $G_i$ such that $|G_j(E)| \leq |G_i(E)|$ for all $j \neq i$. $B_i$ is called the main BAT and is denoted by $B_{\max}$.

The first $2^k$ rows for the first $k$ columns in $B_{\max} = B_{(i)}$ are equal to those in $B_{(k)}$ for all $k \leq i$. All other BATs for subgraphs can be obtained from the main BAT $B_{(i)}$. Table 5 shows the relationships between the main BAT $B_{(4)}$ and sub-BATs $B_{(1)}$, $B_{(2)}$, and $B_{(3)}$.

**Table 5.** The main-BAT $B_{(4)}$ and sub-BAT $B_{(1)}$, $B_{(2)}$, and $B_{(3)}$.

| $i$ | $x_{i,1}$ | $x_{i,2}$ | $x_{i,3}$ | $x_{i,4}$ |
|---|---|---|---|---|
| 1 | 0 | 0 | 0 | 0 |
| 2 | 1 | 0 | 0 | 0 |
| 3 | 0 | 1 | 0 | 0 |
| 4 | 1 | 1 | 0 | 0 |
| 5 | 0 | 0 | 1 | 0 |
| 6 | 1 | 0 | 1 | 0 |
| 7 | 0 | 1 | 1 | 0 |
| 8 | 1 | 1 | 1 | 0 |
| 9 | 0 | 0 | 0 | 1 |
| 10 | 1 | 0 | 0 | 1 |
| 11 | 0 | 1 | 0 | 1 |
| 12 | 1 | 1 | 0 | 1 |
| 13 | 0 | 0 | 1 | 1 |
| 14 | 1 | 0 | 1 | 1 |
| 15 | 0 | 1 | 1 | 1 |
| 16 | 1 | 1 | 1 | 1 |

The pseudocode for finding the main BAT is based on the traditional BAT and is already listed in Section 3.2. Hence, the pseudo-code for finding all vectors for sub-BATs is presented below.

**Algorithm for finding all vectors in a sub-BAT**

**Input:**     The main-BAT $B_{max} = B_{(i)}$.

**Output:**    All feasible $k$-tuple binary-state vectors $X$ in the sub-BAT $B_{(k)}$ for $k = 1, 2, …., (i–1)$.

**STEP S0.** Let $j = 0$.

**STEP S1.** Let $B_{(k),j,h} = B_{(i),j,h}$ for $h = 1, 2, …, k$, where $B_{k,j}(a_h)$ is the element of the $j$th row and the $h$th column of $B_{(k)}$.

**STEP S2.** If $j < k$, let $j = j + 1$ and go to STEP S1.

From the above pseudocode, it is very simple to find all sub-BATs after obtaining the main BAT. The main BAT is found only once and stored in the proposed algorithm. Each sub-BAT is generated from the main is also only once, whereas it is needed in the convolution product, and it is discarded after its convolution product. The convolution product is discussed in Section 5.

For example, $|G_2(E)| = 3 > |G_1(E)| = |G_3(E)| = 2$ from Table 4. Hence, $B_{3,2} = B_2$ for $G_1$ and $B_{3,2} = B_2$ for $G_3$ based on $B_3$ obtained from $G_2$. Thus, $B_{max} = B_3$ is the main BAT in Figure 1 based on Table 4. On the other hand, $B_1$ and $B_2$ are sub-BATs and can be generated from $B_3$.

## 4.4 Source-Target Matrix of Each Vector

Let $\eta$ be the number of self-adjustment subgraphs after using the shortest MCs to split the graphs into disjoint subgraphs $G_1, G_2, …, G_\eta$. From Sections 4.1 and 4.2, each self-adjustment subgraph, say $G_i$ for $i = 1, 2, …., \eta$, has its own source subset $G_i(S)$ and sink subset $G_i(T)$. Note that $G_i(T) = G_{(i+1)}(S)$ for $i = 1, 2, …., (\eta-1)$. In Section 4.3, subgraph $G_i$ also has its own vectors from at least one node in $G_i(S)$ and one node in $G_i(T)$.

To clarify the connections between nodes in $G_i(S)$ and $G_i(T)$ in $X_{i,j}$, a new terminology called the source-target matrix is proposed and defined as follows:

$$M_{i,j} = \begin{bmatrix} m_{1,1} & m_{1,2} & \cdots & m_{1,\sigma} \\ m_{2,1} & m_{2,2} & \cdots & m_{2,\sigma} \\ \vdots & \vdots & \cdots & \vdots \\ m_{\tau,1} & m_{\tau,2} & \cdots & m_{\tau,\sigma} \end{bmatrix} \qquad (7)$$

where $m_{\alpha,\beta} = 1$ denotes that the $\alpha$th source node in $G_i(S)$ is connected to the $\beta$th sink node in $G_i(T)$ for the vector $X_{i,j}$; otherwise, $m_{\alpha,\beta} = 1$. For easy recognition, in the remainder of this study,

$$m_{(\alpha,\beta)} = \begin{cases} 1 & \text{if nodes } \alpha \text{ is connected to } \beta \\ 0 & \text{otherwise} \end{cases}. \qquad (8)$$

$G_1(S) = \{1\}$, $G_\eta(T) = \{n\}$, and $G_i(T) = G_{(i+1)}(S)$ for $i = 1, 2, \ldots, (\eta-1)$. We have the following properties for the source-target matrices:

1. The value of each element in $M_{1,j}$ is either 1 or 0.
2. $M_{1,j}$ is a row for all $j$, e.g., Table 6.
3. $M_{\eta,j}$ is a column for all $j$, e.g., Table 7.
4. The number of columns in $M_{i,j}$ is equal to the number of rows in $M_{(i+1),k}$ for all $j$, $k$, and $i = 1, 2, \ldots, (\eta-1)$.
5. $M_{i,j}$ can be discarded if each element in $M_{i,j} = \mathbf{0}$, that is, there is no connection between any source node and sink node in $X_{i,j}$. For example, $M_{1,1}$ and $M_{3,1}$ can be discarded in Tables 6 and 8, because both are zeros.

Note that each $Pr(X)$ in Tables 6-8 is calculated by assuming that the probabilities of each functioning arc and failed arc are 0.9 and 0.1, respectively, as shown in Figure 1. The source-target matrices can be the same even if their corresponding vectors are different, for example, (1, 1, 0), (1, 0, 1), (0, 1, 1), and (1, 1, 1) are four different vectors, but they all have the same source-target matrix, as shown in Table 7.

**Table 6.** The source-target matrix for $X_{1,i}$ for $i = 1, 2, 3, 4$.

| Old $i$ | New $i$ | $X_{1,i} = (a_1, a_2)$ | $M_{1,i} = \begin{bmatrix} m_{(1,2)} & m_{(1,3)} \end{bmatrix}$ | $Pr(X_{1,i}) = Pr(M_{1,j})$ |
|---|---|---|---|---|
| 1 |   | (0, 0) | $\begin{bmatrix} 0 & 0 \end{bmatrix}$ |   |
| 2 | 1 | (1, 0) | $\begin{bmatrix} 1 & 0 \end{bmatrix}$ | 0.09 |
| 3 | 2 | (0, 1) | $\begin{bmatrix} 0 & 1 \end{bmatrix}$ | 0.09 |
| 4 | 3 | (1, 1) | $\begin{bmatrix} 1 & 1 \end{bmatrix}$ | 0.81 |

**Table 7.** The source-target matrix for $X_{2,j}$ for $j = 1, 2, \ldots, 8$.

| Old $j$ | New $j$ | $X_{2,j} = (a_3, a_4, a_5)$ | $M_{2,j} = \begin{bmatrix} m_{(2,3)} & m_{(2,4)} \\ m_{(3,3)} & m_{(3,4)} \end{bmatrix}$ | $Pr(X_{2,j}) = Pr(M_{2,j})$ |
|---|---|---|---|---|
| 1 |   | (0, 0, 0) | $\begin{bmatrix} 0 & 0 \\ 1 & 0 \end{bmatrix}$ |   |
| 2 | 1 | (1, 0, 0) | $\begin{bmatrix} 1 & 0 \\ 1 & 0 \end{bmatrix}$ | 0.009 |
| 3 | 2 | (0, 1, 0) | $\begin{bmatrix} 0 & 1 \\ 1 & 0 \end{bmatrix}$ | 0.009 |
| 4 | 3 | (1, 1, 0) | $\begin{bmatrix} 1 & 1 \\ 1 & 1 \end{bmatrix}$ | 0.081 |
| 5 | 4 | (0, 0, 1) | $\begin{bmatrix} 0 & 0 \\ 1 & 1 \end{bmatrix}$ | 0.009 |
| 6 | 5 | (1, 0, 1) | $\begin{bmatrix} 1 & 1 \\ 1 & 1 \end{bmatrix}$ | 0.081 |
| 7 | 6 | (0, 1, 1) | $\begin{bmatrix} 1 & 1 \\ 1 & 1 \end{bmatrix}$ | 0.081 |
| 8 | 7 | (1, 1, 1) | $\begin{bmatrix} 1 & 1 \\ 1 & 1 \end{bmatrix}$ | 0.729 |

**Table 8.** The source-target matrix for $X_{3,k}$ for $k = 1, 2, 3, 4$.

| Old $k$ | New $k$ | $X_{3,k} = (a_6, a_7)$ | $M_{3,k} = \begin{bmatrix} m_{(3,5)} \\ m_{(4,5)} \end{bmatrix}$ | $Pr(X_{3,k}) = Pr(M_{3,k})$ |
|---|---|---|---|---|
| 1 |   | (0, 0) | $\begin{bmatrix} 0 \\ 0 \end{bmatrix}$ |   |
| 2 | 1 | (1, 0) | $\begin{bmatrix} 1 \\ 0 \end{bmatrix}$ | 0.09 |
| 3 | 2 | (0, 1) | $\begin{bmatrix} 0 \\ 1 \end{bmatrix}$ | 0.09 |

| 4 | 3 | (1, 1) | $\begin{bmatrix} 1 \\ 1 \end{bmatrix}$ | 0.81 |

**4.5 Aggregated Probability**

From Table 7, we have the same source-target matrix under different probabilities, for example, 0.081 and 0.729 for the last two. To save runtime, the probabilities of these same source-target matrix probabilities are summed together and called the aggregated probability of the related source-target matrix. For example, as shown in the last second row of Table 9, the probability of the source-target matrix is

$$\begin{bmatrix} 1 & 1 \\ 1 & 1 \end{bmatrix} \tag{9}$$

is summed and changed to 0.972.

Table 9. $M_{2,j}$ with aggregated probability $Pr(M_{2,j})$ for $j = 1, 2, …, 5$.

| Old $j$ | New $j$ | $M_{2,j}=\begin{bmatrix} m_{(2,3)} & m_{(2,4)} \\ m_{(3,3)} & m_{(3,4)} \end{bmatrix}$ | $Pr(M_{2,j})$ |
|---|---|---|---|
| 1 | 1 | $\begin{bmatrix} 0 & 0 \\ 1 & 0 \end{bmatrix}$ | 0.001 |
| 2 | 2 | $\begin{bmatrix} 1 & 0 \\ 1 & 0 \end{bmatrix}$ | 0.009 |
| 3 | 3 | $\begin{bmatrix} 0 & 1 \\ 1 & 0 \end{bmatrix}$ | 0.009 |
| 4, 6, 7, 8 | 4 | $\begin{bmatrix} 1 & 1 \\ 1 & 1 \end{bmatrix}$ | 0.081+0.081+0.081+0.729=0.972 |
| 5 | 5 | $\begin{bmatrix} 0 & 0 \\ 1 & 1 \end{bmatrix}$ | 0.009 |

**5. CONVOLUTION PRODUCT**

The convolution product has been implemented in BATs to combine different vectors obtained from different BATs to reduce runtime [41]. However, a new convolution product called the source-target matrix convolution product is proposed based on the source-target matrix rather than the vectors.

## 5.1 Source-Target Matrix Convolution Product

Let vector $Z_i = X_1 \otimes X_2 \otimes \ldots \otimes X_\eta$ and $X_i$ be obtained from $G_i$ from source subset $s_i \subseteq S_i$ to target subset $t_i \subseteq T_i$. Assume that $X(a) = X_i(a)$ for all $a \in G_i(E)$ and $X(a) = X_{(i+1)}(a)$ for all $a \in G_{(i+1)}(E)$, that is, vector $X$ consists of $X_i$ and $X_{(i+1)}$. If $t_i \cap s_{(i+1)} = \emptyset$, then $X$ is disconnected, and there is no need to have more convolution of $X$.

From the above, we can use the source-target matrix to replace the vectors in the convolution product, which is called the source-target matrix convolution product. The convolutional source-target matrix is a new source-target matrix after the source-target matrix convolution product between two source-target matrices.

Based on the 4$^{th}$ property listed in Section 4.4, the source-target matrix convolution product of $M_{i,k_1} \otimes M_{(i+1),i_2}$ is defined and calculated as follows:

$$(M_{i,j} \otimes M_{(i+1),k})_{\alpha,\beta} = \text{Max}\left\{\underset{h}{\text{Min}}\{(M_{i,j})_{\alpha,h}, (M_{(i+1),k})_{h,\beta}\}\right\} \tag{10}$$

For example, the source-target matrix convolution products for the source-target matrices shown in Tables 6 and 9 are listed in Table 10. If $M_{i,j}$ is a row, Eq. (10) can be simplified as follows.

$$(M_{i,j} \otimes M_{(i+1),k})_{1,\beta} = \begin{cases} 1 & (M_{i,j})_{1,h} = (M_{(i+1),k})_{h,\beta} = 0 \text{ for at least one } h \\ 0 & \text{otherwise} \end{cases} \tag{11}$$

**Table 10.** Results of $M_k = M_{1,i} \otimes M_{2,j}$ for $i = 1, 2, 3$ and $j = 1, 2, \ldots, 5$

| k | $M_{1,i}$ | $M_{2,j}$ | $M_{1,i} \otimes M_{2,j} = \begin{bmatrix} m_{(1,3)} & m_{(1,4)} \end{bmatrix}$ | $Pr(M_{1,i} \otimes M_{2,j})$ |
|---|---|---|---|---|
|   | $[1\ 0]$ | $\begin{bmatrix} 0 & 0 \\ 1 & 0 \end{bmatrix}$ | $[0\ 0]$ |   |
| 1 | $[1\ 0]$ | $\begin{bmatrix} 1 & 0 \\ 1 & 0 \end{bmatrix}$ | $[1\ 0]$ | 0.09×0.009=0.00081 |
| 2 | $[1\ 0]$ | $\begin{bmatrix} 0 & 1 \\ 1 & 0 \end{bmatrix}$ | $[0\ 1]$ | 0.09×0.009=0.00081 |
| 3 | $[1\ 0]$ | $\begin{bmatrix} 1 & 1 \\ 1 & 1 \end{bmatrix}$ | $[1\ 1]$ | 0.09×0.972=0.08748 |
|   | $[1\ 0]$ | $\begin{bmatrix} 0 & 0 \\ 1 & 1 \end{bmatrix}$ | $[0\ 0]$ |   |
| 4 | $[0\ 1]$ | $\begin{bmatrix} 0 & 0 \\ 1 & 0 \end{bmatrix}$ | $[1\ 0]$ | 0.09×0.001=0.00009 |
| 5 | $[0\ 1]$ | $\begin{bmatrix} 1 & 0 \\ 1 & 0 \end{bmatrix}$ | $[1\ 0]$ | 0.09×0.009=0.00081 |
| 6 | $[0\ 1]$ | $\begin{bmatrix} 0 & 1 \\ 1 & 0 \end{bmatrix}$ | $[1\ 0]$ | 0.09×0.009=0.00081 |
| 7 | $[0\ 1]$ | $\begin{bmatrix} 1 & 1 \\ 1 & 1 \end{bmatrix}$ | $[1\ 1]$ | 0.09×0.972=0.08748 |
| 8 | $[0\ 1]$ | $\begin{bmatrix} 0 & 0 \\ 1 & 1 \end{bmatrix}$ | $[1\ 1]$ | 0.09×0.009=0.00081 |
| 9 | $[1\ 1]$ | $\begin{bmatrix} 0 & 0 \\ 1 & 0 \end{bmatrix}$ | $[1\ 0]$ | 0.81×0.001=0.00081 |
| 10 | $[1\ 1]$ | $\begin{bmatrix} 1 & 0 \\ 1 & 0 \end{bmatrix}$ | $[1\ 0]$ | 0.81×0.009=0.00729 |
| 11 | $[1\ 1]$ | $\begin{bmatrix} 0 & 1 \\ 1 & 0 \end{bmatrix}$ | $[1\ 1]$ | 0.81×0.009=0.00729 |
| 12 | $[1\ 1]$ | $\begin{bmatrix} 1 & 1 \\ 1 & 1 \end{bmatrix}$ | $[1\ 1]$ | 0.81×0.972=0.7832 |
| 13 | $[1\ 1]$ | $\begin{bmatrix} 0 & 0 \\ 1 & 1 \end{bmatrix}$ | $[1\ 1]$ | 0.81×0.009=0.00729 |

Note that $M_{1,i} \otimes M_{2,j}$ is infeasible and no longer needs to be considered if it is zero, and $\Pr(M_{1,i} \otimes M_{2,j})$ is empty in Table 10.

## 5.2 Consecutive Source-Target Matrix Convolution Product

QB-II combines these results from subproblems using the source-target matrix convolution product in sequence. Let $M_i = M_{1,j_1} \otimes M_{2,j_2} \otimes \cdots \otimes M_{i,j_i} = M_{i-1} \otimes M_{i,j_i}$. Because there is only one source node, that is, node 1, $M_i$ is a row, and we have:

$$(M_{i-1} \otimes M_{i,j_i})_{1,\beta} = \begin{cases} 1 & (M_{i-1})_{1,h} = (M_{i,j_i})_{h,\beta} = 1 \text{ for at least one } h \\ 0 & \text{otherwise} \end{cases} \quad (12)$$

based on Eq. (11).

For example, each source-target matrix shown in Table 10 is also the result of the consecutive source-target matrix convolution product from $G_1$ (see Table 6) to $G_2$ (see Table 9).

## 5.3 Aggregated Probability After Source-Target Matrix Convolution Product

Because $\Pr(X_{i,j} \otimes X_{(i+1),k}) = \Pr(X_{i,j}) \Pr(X_{(i+1),k})$, if $X_{i,j} \otimes X_{(i+1),k}$ is connected. Hence, $\Pr(M_{i,j} \otimes M_{(i+1),k}) = \Pr(M_{i,k_1}) \Pr(M_{(i+1),i_2})$ if $M_{i,k_1} \otimes M_{(i+1),i_2}$ is not a zero matrix. For example, each value of $\Pr(M_{1,i} \otimes M_{2,j})$ is shown in the last column of Table 10, if $M_{1,i} \otimes M_{2,j}$ is feasible.

Let $M_i = M_{1,j_1} \otimes M_{2,j_2} \otimes \cdots \otimes M_{i,j_i} = M_{i-1} \otimes M_{i,j_i}$. In the same way, we have

$$\Pr(M_i) = \Pr(M_{1,j_1}) \times \Pr(M_{2,j_2}) \times \cdots \times \Pr(M_{i,j_i}) = \Pr(M_{i-1}) \times \Pr(M_{i,j_i}) \qquad (13)$$

Similar to the aggregated probability of the source-target matrix discussed in Section 4.5, the probabilities of the same source-target matrix after the source-target matrix convolution product can be added together to reduce the number of convolutional source-target matrices. For example, Table 10 is simplified by aggregating the probability of the convolutional source-target matrix [1, 0] and [1, 1], as shown in the 2$^{nd}$ and the 3$^{rd}$ columns in Table 11. Subsequently, the number of convolutional source-target matrices is reduced from 13 to 3. Note that the number of convolutional source-target matrices is 22 if the aggregated probability of both the source-target matrix and convolutional source-target matrices are not implemented.

Table 11. Aggregated results of $M_{1,i} \otimes M_{2,j}$ for $i = 1, 2, 3$ and $j = 1, 2, \ldots, 8$

| $i$ | $j$ | $\begin{bmatrix} m_{(1,3)} & m_{(1,4)} \end{bmatrix}$ | $\sum_{i,j} M_{1,i} \otimes M_{2,j}$ |
|---|---|---|---|
| 1, 4, 5, 6, 9, 10 | 1 | $\begin{bmatrix} 1 & 0 \end{bmatrix}$ | 0.00081+0.00009+0.00081+ 0.00081+0.00081+0.00729=0.01062 |
| 2 | 2 | $\begin{bmatrix} 0 & 1 \end{bmatrix}$ | 0.00081 |
| 3, 7, 8, 11, 12, 13 | 3 | $\begin{bmatrix} 1 & 1 \end{bmatrix}$ | 0.08848+0.08848+0.00081+ 0.00729+0.7832+0.00729=0.97767 |

## 6. PROPOSED QB-II

The pseudocode of the proposed QB-II is presented in Section 6.1, based on Sections 4 and 5.

An example is provided to demonstrate the proposed algorithm in Section 6.2. Section 6.3 conducts experiments on 20 benchmark problems to have a complete comparison among the QB-II and the quick BAT which is the best-known algorithm currently. In addition, the effectiveness of the number of subgraphs was tested to validate the concepts of the proposed main BAT and sub-BATs.

### 6.1 Pseudo-Code

The pseudocode of the proposed QB-II is given below.

**Algorithm for the proposed QB-II**

**Input:** A binary-state network $G(V, E, \mathbf{D})$ with the source node 1 and the sink node $n = |V|$.

**Output:** $R(G)$.

**STEP 0.** Find all shortest MCs, say $C_1, C_2, \ldots, C_{\eta-1}$, where $C_k$ separates $G_k$ and $G_{k+1}$ for $k = 1, 2, \ldots, \eta-1$, and self-adjustment subgraphs based on Section 4.2.

**STEP 1.** Find $B_{\max}$, where $|E_j| \leq |E_{\max}|$ for all $j = 1, 2, \ldots, \eta$, and let $i = 1$.

**STEP 2.** Let $B_{(g)}$ be the set of the first $2^g$ $g$-tuple vectors in $B_{\max}$, where $g = |E_i|$.

**STEP 3.** Find and calculate the aggregated probability of the source-target matrix for each vector in $B_{(g)}$ and let $K_i$ be number of the aggregated source-target matrices related to $B_{(g)}$.

**STEP 4.** If $i = 1$, let $\Pi = \{\Pi_j = M_{1,j} \mid j = 1, 2, \ldots, K_1\}$, $i = i + 1$, $K = K_1$, and go to STEP 2.

**STEP 5.** Let $k^* = h = j = 1$.

**STEP 6.** If $\Pi^*_{k^*} = \Pi_h \otimes M_{i,j}$ is not zero, calculate $\Pr(\Pi^*_{k^*})$ and let $k^* = k^* + 1$.

**STEP 7.** If $j < K_i$, let $j = j + 1$ and go to STEP 6.

**STEP 8.** If $h < K$, let $h = h + 1, j = 1$, and go to STEP 6.

**STEP 9.** If $i < \eta$, let $\Pi = \{\Pi_k = \Pi^*_k \mid k = 1, 2, \ldots, k^*\}$, $K = |\Pi|$ after aggregating elements in $\Pi$, and go to STEP 2.

**STEP 10.** The final aggregate probability is the reliability.

Based on the procedure proposed in Section 4.1, STEP 0 finds all the shortest MCs to separate $G(V, E)$ into $G_1 = G(V_1, E_1)$, $G_2 = G(V_2, E_2)$, ..., and $G_\eta = G(V_\eta, E_\eta)$, such that $E_i \cap E_j = \emptyset$. STEP 1 finds the main BAT, and the rest of the sub-BAT is generated in STEP 2 based on STEP 4.3.

STEP 3 finds aggregated source-target matrices for connected vectors together with their probability. The loops from STEPs 5 to 9 implement the proposed source-target matrix convolution product to find feasible aggregated convolutional source-target matrices and their probabilities. STEP10 outputs the final reliability of the binary-state network.

Each connected vector has only one feasible source-target matrix. Hence, the number of aggregated source-target matrices is less than or equal to that of the connected super vectors in the quick BAT.

The major time complexity for QB-II is the number of source-target matrices. Each source node can also be a target node in a subgraph, for example, node 3 in the 2$^{nd}$ subgraph of Figure 1. The number of source-target matrices is $O(2^{2n})$. The second part of the time complexity is the source-target matrix convolution product between $\Pi_h$ and $M_{i,j}$ in STEP 6. Matrix $\Pi_h$ is a row, and its element number is equal to the number of columns in $M_{i,j}$, that is, the time complexity is $O(n^2)$ for two source-target matrices. Hence, the time complexity of QB-II is $O(n^2 2^{2n})$ in worst cases.

## 6.2 Example

It is difficult to calculate the exact binary-state network reliability, which is both an NP-hard problem and a #P-Hard problem [9, 10, 11]. The worst part is that the difficulty increases exponentially with problem size. Hence, it is not always appropriate to exemplify a new algorithm using large-scale examples. Similar to the quick BAT, an intermediate example (Figure 5) is employed to demonstrate the proposed algorithm step-by-step to allow readers to understand the proposed QB-II quickly as follows.

**STEP 0.** From Section 4.1, there is only one shortest path: $\{a_1, a_6\}$, we have two shortest MCs: $C_1 = \{a_1, a_2\}$ and $C_2 = \{a_6, a_7\}$, i.e., $k = 3$. $C_1$ and $C_2$ separate $G(V, E)$ into $G_1 = (V_1 =\{1\}, V_1 =\{a_1, a_2\})$, $G_2 = (V_2 =\{2, 3, 4\}, E_2 =\{a_3, a_4, a_5\})$, and $G_3 = (V_3 =\{5\}, E_2 =\{a_6, a_7\})$ after self-adjustment to subgraphs.

**STEP 1.** Find $B_{max} = B_{(3)} = B_2$ because both $|E_1| = |E_3| = 2$ are less than $|E_2| = 3$ and let $i = 1$.

**STEP 2.** Because $|E_1| = 2$, find $B_{(2)}$ from $B_{max}$ as shown in $X_{1,i} = (a_1, a_2)$ of Table 6.

**STEP 3.** Find each source-target matrix and calculate its probability as shown in $M_{1,i} = \begin{bmatrix} m_{(1,2)} & m_{(1,3)} \end{bmatrix}$ and $Pr(X_{1,i})=Pr(M_{1,j})$ of Table 6, respectively, and let $K_1 = 3$.

**STEP 4.** Because $i = 1$, let $\Pi = \{M_{1,1}, M_{1,2}, M_{1,3}\}$, $i = i + 1 = 2$, $K = K_1 = 3$, and go to STEP 2.

**STEP 2.** Because $|E_2| = 3$, $B_{(3)} = B_{max}$ as shown in Table 7.

**STEP 3.** Find each aggregated source-target matrix and calculate its probability as shown in Table 9, respectively, and let $K_2 = 7$.

**STEP 4.** Because $i > 1$, go to STEP 5.

**STEP 5.** Let $k^* = h = j = 1$. Note that

**STEP 6.** Because $\Pi_1^* = \Pi_1 \otimes M_{2,1} = \begin{bmatrix} 1 & 0 \end{bmatrix} \otimes \begin{bmatrix} 0 & 0 \\ 1 & 0 \end{bmatrix} = \mathbf{0}$, go to STEP 7.

**STEP 7.** Because $j = 1 < K_2 = 7$, let $j = j + 1 = 2$ and go to STEP 6.

**STEP 6.** Because $\Pi_1^* = \Pi_1 \otimes M_{2,2} = \begin{bmatrix} 1 & 0 \end{bmatrix} \otimes \begin{bmatrix} 1 & 0 \\ 1 & 0 \end{bmatrix} = \begin{bmatrix} 1 & 0 \end{bmatrix}$, calculate $Pr(\Pi_1^*) = 0.09 \times 0.009 = 0.00081$, and let $k^* = k^* + 1 = 2$.

$$\vdots$$

**STEP 9.** Because $i = 2 < \eta = 3$, let $\Pi = \{\Pi_1 = \begin{bmatrix} 1 & 0 \end{bmatrix}, \Pi_2 = \begin{bmatrix} 0 & 1 \end{bmatrix}, \Pi_3 = \begin{bmatrix} 1 & 1 \end{bmatrix}\}$ after aggregating elements in $\Pi^*$, $K = |\Pi| = 3$, calculate $Pr(\Pi_k)$ for $k = 1, 2, 3$, as shown in Table 11, and go to STEP 2.

$$\vdots$$

**STEP 9.** Because $i = \eta = 3$, go to STEP 10.

**STEP 10.** The final aggregate probability 0.97818 is the reliability as shown in Table 13.

**Table 12.** Results of $M_h \otimes M_{3,k}$ for $h$ = 1, 2, 3 and $k$ = 1, 2, 3

| $h$ | $k$ | $M_h$ | $M_{3,k}$ | $M_h \otimes M_{3,k} = [m_{(1,5)}]$ | $Pr(M_h \otimes M_{3,k})$ |
|---|---|---|---|---|---|
| 1 | 1 | $[1\ 0]$ | $\begin{bmatrix}1\\0\end{bmatrix}$ | [1] | 0.01062×0.09=0.0009558 |
|   | 2 | $[1\ 0]$ | $\begin{bmatrix}0\\1\end{bmatrix}$ |  |  |
|   | 3 | $[1\ 0]$ | $\begin{bmatrix}1\\1\end{bmatrix}$ | [1] | 0.01062×0.81=0.0086022 |
| 2 | 4 | $[0\ 1]$ | $\begin{bmatrix}1\\0\end{bmatrix}$ |  |  |
|   | 5 | $[0\ 1]$ | $\begin{bmatrix}0\\1\end{bmatrix}$ | [1] | 0.00081×0.09=0.0000729 |
|   | 6 | $[0\ 1]$ | $\begin{bmatrix}1\\1\end{bmatrix}$ | [1] | 0.00081×0.81=0.000656 |
| 3 | 7 | $[1\ 1]$ | $\begin{bmatrix}1\\0\end{bmatrix}$ | [1] | 0.97767×0.09=0.0879903 |
|   | 8 | $[1\ 1]$ | $\begin{bmatrix}0\\1\end{bmatrix}$ | [1] | 0.97767×0.09=0.0879903 |
|   | 9 | $[1\ 1]$ | $\begin{bmatrix}1\\1\end{bmatrix}$ | [1] | 0.97767×0.81=0.791913 |
| SUM |  |  |  |  | 0.97818 |

In the proposed QB-II for Figure 1, there are 15 aggregated source-target matrices, that is, 3, 5, 3, 3, and 1 for $M_{1,i}$, $M_{2,j}$, $M_{1,i} \otimes M_{2,j}$, $M_{3,k}$, $M_h \otimes M_{3,k}$, respectively. In addition, there were 15 aggregated probabilities. The comparisons between the quick BAT and QB-II tested in Figure 1 are listed in Table 14.

**Table 13.** Comparisons between the quick BAT and QB-II on Figure 1.

|  | vectors[*] | variables | connections | multiplications | summations |
|---|---|---|---|---|---|
| QB-II | **11** | **81**[!] | 4+8+4=**16** | **47**[#] | **15** |
| quick BAT | 26 | 165 | 26 | 165 | 26 |

[*] The number of supervectors in the quick BAT and target-source matrices in QB-II.
[!] 4×2+8×4+4×2+13×2+7×1 = 81
[#] 3×2+5×3+3×2+13+7 = 47

From Table 14, it can be seen that QB-II is more efficient than the quick BAT.

## 6.3 Experiments

The proposed QB-II is compared with the quick BAT and BDD on 20 benchmark binary-state networks (Figure 7(1) – (20) [39, 40, 41, 4, 30]). We only compared QB-II, quick BAT, and BDD because the quick BAT already outperformed DFS, for example, the quick inclusion-exclusion method (QIE) [44], and BFS, for example, the RSDP [30].

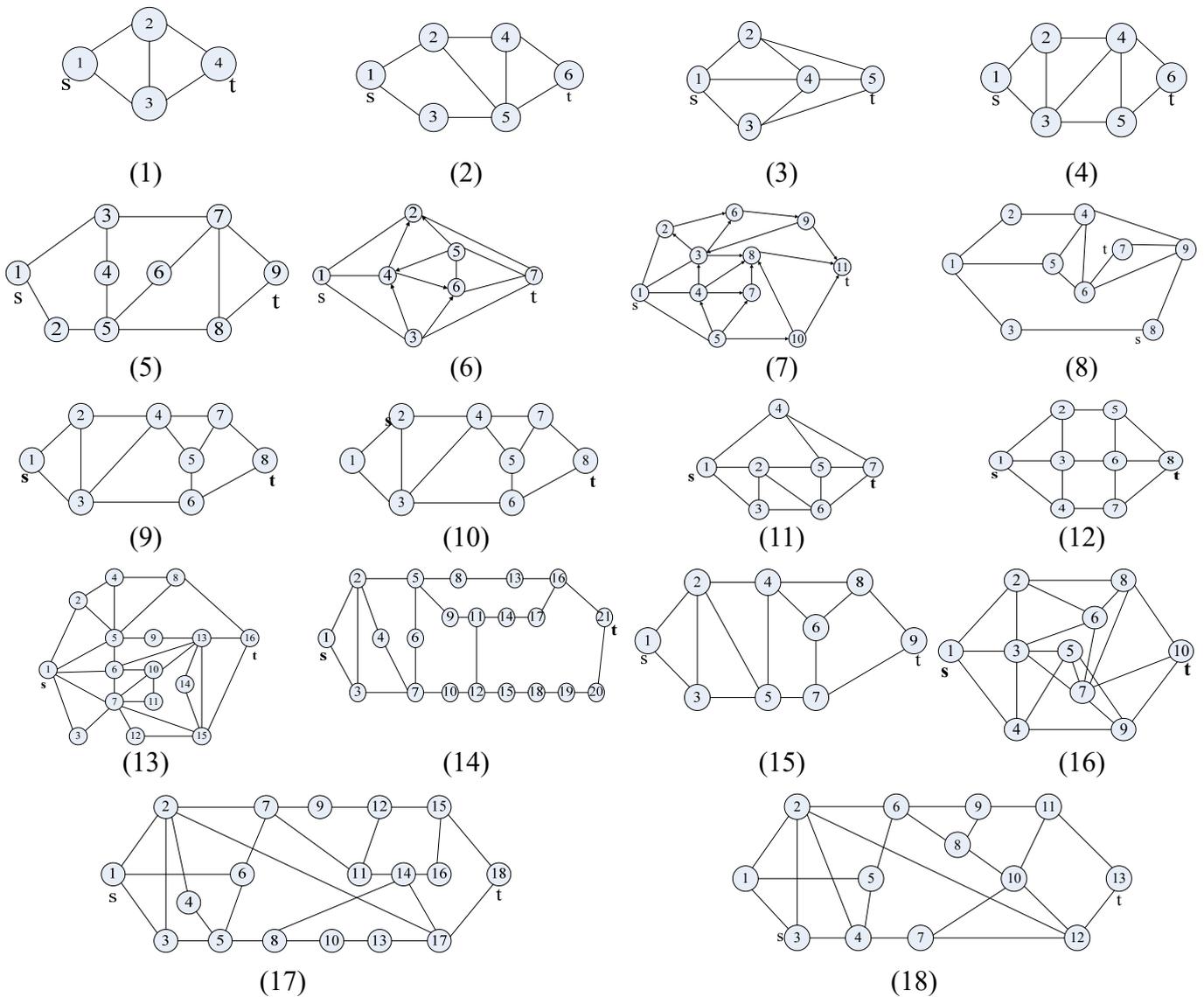

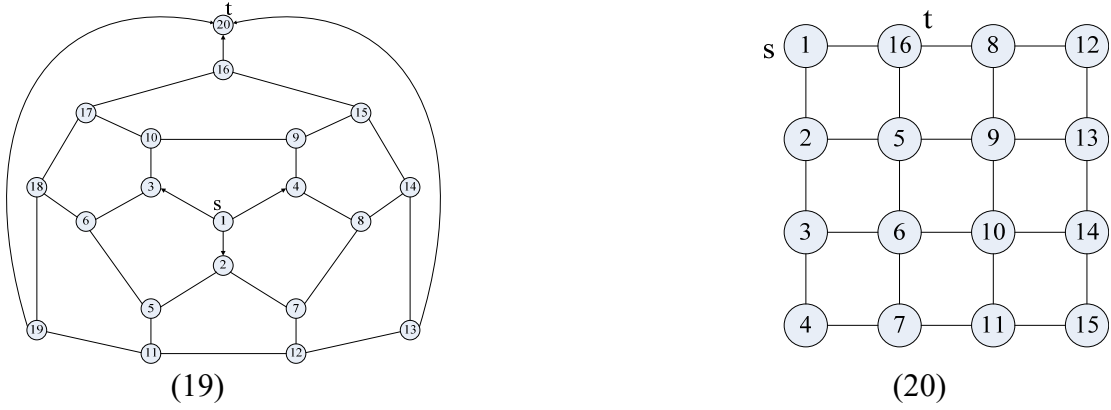

(19)  (20)

**Figure 7.** 20 benchmark binary-state networks used in the test

The test environments for the comparisons were set to those reported in [44], as listed in Table 15, to ensure a fair evaluation.

**Table 14.** Computer Environments and problem settings.

| Item | Environment |
|---|---|
| Operation System | 64-bit Windows 10 |
| Compiler | DEV C$^{++}$ 5.11 |
| CPU | Intel Core i7-6650U @ 2.20GHz 2.21GHz |
| RAM | 16 GB |
| Stopping Criteria | 10 hours |
| Arc reliability | 0.9 |

The notations $T_{QB\text{-}II}$, $T_{QB}$, and $T_{BDD}$ represent the runtimes obtained from the proposed QB-II, quick BAT, and BDD, respectively. The runtime results of both the quick BAT and BDD were obtained from [40]. The QB-II, quick BAT, and BDD are based on different special "solutions," i.e., the source-target matrix, the super vector, and the connected vector, respectively. To obtain a fair comparison easily and straightforwardly, we only focus on the runtime of each algorithm. The best values are shown in bold font.

**Table 15.** Comparison of the QB-II, quick BAT, and BDD.

| Figure 7 | $R$ | $T_{QB\text{-}II}$ | $T_{QBAT}$ | $T_{BDD}$ |
|---|---|---|---|---|
| (1) | 0.9784800000 | **0** | 0 | 0 |
| (2) | 0.9684254700 | **0** | 0 | 0 |
| (3) | 0.9976316400 | **0** | 0 | 0 |
| (4) | 0.9771844050 | **0** | 0 | 0 |
| (5) | 0.9648551232 | **0** | 0 | 0 |
| (6) | 0.9966644040 | **0** | 0 | 0 |
| (7) | 0.9940757879 | 0.0179686 | 0.139 | 0.151 |
| (8) | 0.9691117946 | 0 | 0 | 0 |
| (9) | 0.9751158974 | 0 | 0 | 0 |

| (10) | 0.9840681530 | 0 | 0 | 0.001 |
| (11) | 0.9974936737 | 0 | 0 | 0 |
| (12) | 0.9962174933 | 0 | 0 | 0 |
| (13) | 0.9988659386 | 187.1002947 | **178.292** | 194.657 |
| (14) | 0.9045770823 | **0.008965884** | 5.074 | 6.951 |
| (15) | 0.9741454748 | **0** | **0** | **0** |
| (16) | 0.9975059058 | **0** | **0** | 0.002 |
| (17) | 0.9858583145 | **2.05686E-05** | 16.392 | 18.839 |
| (18) | 0.9873899673 | **1.16354E-06** | 0.455 | 0.536 |
| (19) | 0.9971203988 | **1.000173957** | 131.040 | 147.669 |
| (20) | 0.9878311486 | **1.75236E-05** | 0.731 | 1.067 |

All three algorithms can solve Figures 7(1) – (12), (15), (16), (18), and (20) within one minute; Figures 7(14) and (17) within 10 min; and the rest within four hours. In general, QB-II is more efficient than quick BAT and BDD. There are two interesting phenomena: QB-II can solve Figure 7(19) in almost one minute but the other two took more than two hours for that. Additionally, QB-II is 10 min slightly less efficient, as shown in Figure 7(13), than quick BAT. The two main reasons for these phenomena are explained below.

1. QB-II is a node-based algorithm rather than an arc-based one.

    As mentioned before, both the quick BAT and BDD count on the super vectors and connected vectors, respectively. Both super vectors and connected vectors are vectors in arcs; for example, the value of the first coordinate of the vector is the state of the first arc. Hence, quick BAT and BDD are generally arc-based algorithms.

    In contrast, the proposed QB-II transfers each arc-based subvector to a node-based source-target matrix in the beginning, and these node-based source-target matrices are throughout the QB-II process. Because $O(n^2) = O(m)$, in general, the number of nodes $n$ is equal to that of arcs $m$. Hence, the node-based algorithm is more efficient than the arc-based algorithm, as shown in most cases in Figure 7.

    Moreover, from the discussion of the QB-II time complexity, if $(2n)$ is close to $m$, the arc-based algorithm has a better chance of beating the proposed QB-II, as shown in Figure 7(13).

2. A smart divide-and-conquer strategy.

As shown in [41], the BAT-based algorithm can beat the best AI in solving redundancy allocation problems (RAPs) by dividing the series-parallel RAP into subproblems and intelligently solving and connecting these subproblems. Analogous to the above situations, QB-II divides the entire problem into subproblems and connects these subproblems in series using the new convolution product and aggregation intelligently.

From the above fair comparisons, the QB-II improves the efficiency of the quick BAT and is more attractive than the quick BAT and BDD for these problems with $(2n) \ll m$. Note that the performance of the quick BAT is better than most binary-state network reliability problems such as the BAT [39], QIE [44], and RSDP [30].

# 7. CONCLUSIONS

Network reliability is an important index for evaluating network performance because it reflects the probability of the current state of networks. In this study, a new algorithm called QB-II is proposed to improve the efficiency of the quick BAT in calculating the reliability of a binary-state network, which is fundamental to all types of networks.

The proposed QB-II presented novel concepts, including the shortest MCs to separate the original BAT into main-BAT and sub-BATs, source-target matrices to replace the vectors, new convolution products to connect source-target matrices from different BATs, and aggregation to reduce the number of probability calculations and connectivity verifications. From computational experiments conducted on 20 benchmark problems, the performance of QB-II is confirmed by comparing it with the quick BAT.

The proposed QB-II will be integrated with approximated-reliability algorithms, for example, Monte Carlo simulation, to improve its efficiency for larger-sized problems without sacrificing too much solution quality. QB-II will also be reinforced from the binary-state reliability problem to solve more real-life applications in future works.

**ACKNOWLEDGEMENTS**

This research was supported in part by the Ministry of Science and Technology, R.O.C. under grant MOST 107-2221-E-007-072-MY3 and MOST 110-2221-E-007-107-MY3. This article was once submitted to arXiv as a temporary submission that was just for reference and did not provide the copyright.

[8]  M. Wang, W.C. Yeh, T.C. Chu, X. Zhang, C.L. Huang, and J. Yang, "Solving Multi-Objective Fuzzy Optimization in Wireless Smart Sensor Networks under Uncertainty Using a Hybrid of IFR and SSO Algorithm," *Energies*, https://doi.org/10.3390/en11092385, 2018.

[9]  D. Shier, *Network Reliability and Algebraic Structures*, Clarendon Press, New York, NY, USA, 1991.

[10] C. J. Colbourn, *The combinatorics of network reliability*, Oxford University Press, 1987.

[11] G. Levitin, *The universal generating function in reliability analysis and optimization*, Springer, 2005.

[12] Á. Rodríguez-Sanz, D. Á. Álvarez, F. G. Comendador, R. A. Valdés, J. Pérez-Castán, and M. N. Godoy, "Air Traffic Management based on 4D Trajectories: A Reliability Analysis using Multi-State Systems Theory," *Transportation research procedia,* vol. 33, pp. 355-362, 2018.

[13] P. Wang, R. Billinton, and L. Goel, "Unreliability cost assessment of an electric power system using reliability network equivalent approaches," *IEEE Transactions on power systems,* vol. 17, no. 3, pp. 549-556, 2002.

[14] J. E. Ramirez-Marquez, "Assessment of the transition-rates importance of Markovian systems at steady state using the unscented transformation," *Reliability Engineering & System Safety*, vol. 142, pp. 212-220, 2015.

[15] S. Laitrakun and E. J. Coyle, "Reliability-based splitting algorithms for time-constrained distributed detection in random-access WSNs," *IEEE Transactions on Signal Processing,* vol. 62, no. 21, pp. 5536-5551, 2014.

[16] C. M. R. Sanseverino and J. E. Ramirez-Marquez, "Uncertainty propagation and sensitivity analysis in system reliability assessment via unscented transformation," *Reliability Engineering & System Safety*, vol. 132, pp. 176-185, 2014.

[17] C. L. Huang, "A particle-based simplified swarm optimization algorithm for reliability redundancy allocation problems," *Reliability Engineering & System Safety*, vol. 142, pp. 221-230, 2015.

[18] W. C. Yeh, "An Improved Method for the Multistate Flow Network Reliability with Unreliable Nodes and the Budget Constraint Based on Path Set", *IEEE Transactions on Systems*, *Man*, *and Cybernetics: Systems* (renamed from *IEEE Transactions on Systems, Man, and Cybernetics -- Part A: Systems and Humans*), vol. 41, no. 2, pp. 350-355, 2011.

[19] S. Chakraborty, N. K. Goyal, S. Mahapatra, and S. Soh, "Minimal Path-Based Reliability Model for Wireless Sensor Networks With Multistate Nodes," *IEEE Transactions on Reliability,* vol. 69, no. 1, pp. 382-400, 2019.

[20] W. C. Yeh, C. Bae, and C. L. Huang, "A new cut-based algorithm for the multi-state flow
- 34 -